\documentclass[journal]{IEEEtran}
\usepackage{amsmath,amsfonts}
\usepackage{graphicx,amssymb}
\usepackage{algorithmic,algorithm}
\usepackage{amsthm,multirow}
\usepackage{subfigure,url,color,capt-of,booktabs,hhline}
\usepackage[noadjust]{cite}

\usepackage{cite}

\usepackage{graphics}

\ifCLASSINFOpdf
\else
\fi
\begin{document}
%
\title{Interference Mitigation for Network-Level ISAC: An Optimization Perspective}
%
%
%

\author{Dongfang Xu, Yiming Xu, Xin Zhang, Xianghao Yu, Shenghui Song, and Robert Schober
\thanks{Dongfang Xu, Yiming Xu, Xin Zhang, and Shenghui Song are with the Hong Kong University of Science and Technology, Hong Kong, China; 
Xianghao Yu is with City University of Hong Kong, Hong Kong, China;
Robert Schober is with Friedrich-Alexander University Erlangen-Nuremberg, Germany.}}
\maketitle

\begin{abstract}
Future wireless networks are envisioned to simultaneously provide high data-rate communication and ubiquitous environment-aware services for numerous users. One promising approach to meet this demand is to employ network-level integrated sensing and communications (ISAC) by jointly designing the signal processing and resource allocation over the entire network. However, to unleash the full potential of network-level ISAC, some critical challenges must be tackled. Among them, interference management is one of the most significant ones. In this article, we build up a bridge between interference mitigation techniques and the corresponding optimization methods, which facilitates efficient interference mitigation in network-level ISAC systems. In particular, we first identify several types of interference in network-level ISAC systems, including self-interference, mutual interference, crosstalk, clutter, and multiuser interference. Then, we present several promising techniques that can be utilized to suppress specific types of interference. For each type of interference, we discuss the corresponding problem formulation and identify the associated optimization methods. Moreover, to illustrate the effectiveness of the proposed interference mitigation techniques, two concrete network-level ISAC systems, namely coordinated cellular network-based and distributed antenna-based ISAC systems, are investigated from interference management perspective. Experiment results indicate that it is beneficial to collaboratively employ different interference mitigation techniques and leverage the network structure to achieve the full potential of network-level ISAC. Finally, we highlight several promising future research directions for the design of ISAC systems. 
\end{abstract}



%
\IEEEpeerreviewmaketitle

\section{Introduction}
\begin{figure*}[t]
\centering\includegraphics[width=4.8in]{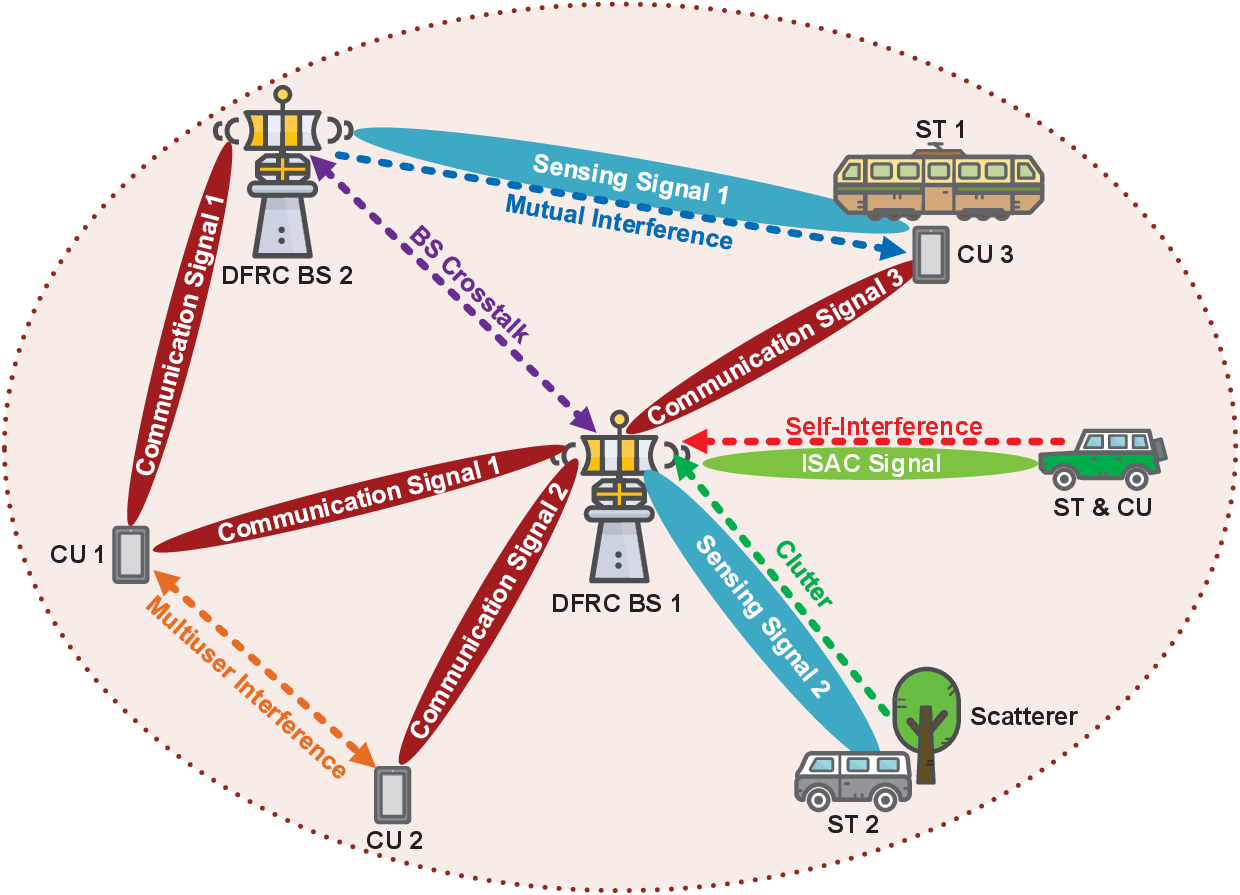}
\caption{Illustration of a network-level ISAC system where two DFRC BSs serve three CUs, two STs, and one ISAC user. The colored ellipses represent the transmit signals while the dashed-line arrows denote the interference in the ISAC system. The communication signal carries the information symbols for the CUs and the sensing signal is employed to sense the desired target. The ISAC signal refers to a signal that can provide sensing and communication services simultaneously.}\label{Fig1_Interference}
\end{figure*}
Wireless networks have been playing a very critical role in connecting people and the world. Looking ahead, wireless networks are envisioned to offer various location-aware services such as smart transportation, virtual/augmented reality, and environmental monitoring \cite{9376324}. To support these fascinating applications, the sixth-generation (6G) networks are envisaged to offer accurate sensing capability. To this end, some early works proposed an orthogonal scheme where radar and communication systems coexist in the same radio propagation environment but operate in different frequency bands. Although the orthogonal paradigm can suppress the interference between sensing and communication, it degrades the power/spectrum efficiency and leads to inefficient utilization of hardware. To circumvent these issues, integrated sensing and communication (ISAC) has emerged as a pivotal enabler in providing sensing services in the 6G networks \cite{9737357}. In a nutshell, ISAC enables sensing and communication in one system. There are four levels of integration between sensing and communication within the ISAC framework, i.e., spectrum-level, hardware-level, signal-level, and network-level integration \cite{8999605,9705498}. For spectrum-level and hardware-level integration, sensing and communication systems share the frequency band and the hardware components, respectively. By signal-level integration, the signal processing scheme and resource allocation policy of a node, e.g., a base station (BS) or an access point (AP), are jointly designed to achieve ISAC. Building upon the first three levels of integration, a more comprehensive and promising architecture, i.e., network-level ISAC, can be realized by jointly designing the signal processing and resource allocation for the entire ISAC network in a more efficient manner.
\par
There are several advantages of employing network-level ISAC, including networked sensing and sensing-assisted communication. However, significant challenges must be tackled before we can exploit the full potential of network-level ISAC systems. Among them, interference mitigation is one of the most difficult ones to handle, due to several different types of interference. First, typical interference for radar and communication systems, e.g., clutter and multiuser interference (MUI), will occur in ISAC systems. Second, the amalgamation nature of ISAC systems causes unprecedented interference. This includes the self-interference (SI) due to the simultaneous transmission and reception of communication and sensing signals, and the inherent mutual interference (MI) between sensing and communication. Third, the network-level integration introduces additional crosstalk between different nodes in ISAC systems. Hence, compared with conventional wireless networks that mainly focus on suppressing MUI, the interference mitigation problem in ISAC systems is not only unprecedented but also more serious. Therefore, it is necessary to develop more advanced and ISAC-specific interference mitigation techniques, together with effective optimization algorithms. 
\par
Several early works, e.g., \cite{8999605,9705498,10012421}, have briefly overviewed the interference issue in ISAC systems and proposed a few promising interference mitigation techniques. However, the above works aimed to sketch the blueprint of ISAC systems and the interference mitigation in network ISAC systems was not their focus. In particular, these works did not cover the origins of each interference in detail and a systematic overview of the dedicated mitigation techniques for each interference is missing. Moreover, the corresponding optimization problems for interference mitigation and suitable methods for solving the corresponding optimization problems have not been well investigated. Among the works that are highly related to this paper, the authors of \cite{9842350,10273396} investigated cooperative interference mitigation for network-level ISAC systems. However, these works only focused on specific interferences, namely SI and MI, and ignored other key interferences in network-level ISAC systems, such as clutter and crosstalk, which leaves a gap in the interference suppression of network-level ISACs.
\par
Focusing on network-level ISAC systems, this paper aims to provide a comprehensive and systematic overview of various interference, the corresponding mitigation techniques, and associated optimization methods. The objective is to build up a bridge between the interference mitigation techniques and the corresponding optimization methods, which serves as a guideline for the design of network-level ISAC. From an optimization perspective, we aim to identify the optimization problem for given interference mitigation methods and suitable optimization methods for the formulated interference mitigation problem. To the best of the authors' knowledge, this is the first article that investigates the interference mitigation issue in network-level ISAC systems. In particular, we first identify the relevant types of interference in ISAC systems including SI, MI, clutter, crosstalk, and MUI. Then, we propose several promising techniques to combat these interferences. Finally, we demonstrate the formulation of the optimization problems for each interference mitigation technique and propose the corresponding optimization algorithms. To provide a concrete understanding of the interference mitigation techniques and the corresponding optimization methods, we apply the proposed techniques for interference mitigation in two specific network-level ISAC systems. In addition, interesting open problems and promising future research topics are highlighted.
\section{Types of Interference in ISAC Systems}\label{Sec:IRSmodel}
In this section, we identify several important types of interference that exist in network-level ISAC systems. In particular, we clarify the origination of each interference and discuss their impacts on the system performance. An exemplary network-level ISAC system is illustrated in Fig. \ref{Fig1_Interference}, where several typical types of interference are depicted.
\begin{itemize}
        \item \textbf{Self-Interference}: Conventional wireless communication systems usually focus on one-way and single-hop transmission and reception. In contrast, radar systems have to deal with round-trip links as the reflected signals carry the desired sensing information. As a result, when implementing ISAC in wireless communication networks, dual-functional radar and communication (DFRC) base stations (BSs) may receive the echo signal before completing the signal transmission. In fact, if the distance between the DFRC BS and a sensing target (ST) is $300$ meters, the round-trip delay of the echo signal is only $2$ $\mu$s. This is much shorter than the $10$ ms symbol duration in a typical frame of the fifth generation new radio \cite{access2020newradio}. This inevitably results in SI between the signal transmission and echo signal reception. Moreover, for the aforementioned scenario, the corresponding round-trip path loss for free-space propagation is around $180$ dB for a sensing signal with carrier frequency 2.4 GHz. However, conventional SI cancellation methods developed for full-duplex communication systems can only provide SI cancellation on the order of $100$ dB, which is not sufficient to effectively mitigate the SI in ISAC systems \cite{bharadia2013full}.
	\item \textbf{Mutual Interference}: Simultaneous sensing and 
        communication is enabled by the employment of ISAC signals. Depending on the application scenario, ISAC signals can be constructed in two different ways. Firstly, dedicated communication signals and sensing signals can be generated separately and mixed in the radio frequency chain. Yet, this inevitably introduces MI between sensing and communication. On the one hand, a dedicated sensing signal causes interference for the communication users (CUs). On the other hand, the communication signals may be reflected by the numerous scatterers in the channel and this may increase the noise level at the sensing receiver, which potentially impairs the detection of the echo signals. Secondly, the ISAC signal can be generated by embedding the information symbol into the sensing beam. In this case, although the joint ISAC signal itself will not cause interference to the CUs, it may still cause additional interference. In particular, to effectively combat the round-trip path loss and conduct reliable sensing, the power of the joint ISAC signal is usually significantly higher than that of conventional communication signals. As a result, the CUs may receive a delayed strong ISAC signal reflected by scatterers (e.g., the ST). In general, the channel state information (CSI) between the CUs and the ST is difficult to obtain as the BS only transmits pilot signals to the CUs. Hence, the delayed strong ISAC signal becomes a sensing-induced interference and cannot be neglected at the receiver side, which can severely jeopardize the information reception of the CUs.
        \item \textbf{Clutter}: In conventional radar systems, sensing signals rebounded by non-interested mobile targets or fixed environmental scatterers cause unfavorable clutter at the receiver side \cite{skolnik2008radar}. In ISAC systems, this problem is exacerbated because, in typical terrestrial scenarios, the desired ST is surrounded by a large number of scatterers. Hence, clutters originating from different directions merge at the DFRC BS and elevate the noise floor, which impedes the detection of echo signals.
        \item \textbf{Crosstalk}: Distributed network architectures are widely adopted in both wireless communication systems (e.g., distributed antenna systems and relay systems) and radar systems (e.g., bistatic radar systems and multistatic radar systems). For network-level ISAC, we need to carefully deal with the crosstalk between different DFRC BSs. Specifically, crosstalk occurs when one DFRC BS receives signals transmitted by another DFRC BS. Even for bi-statistic radar systems, where sensing signal transmission and reception are performed by different DFRC BSs, crosstalk still exists \cite{8999605}. In particular, DFRC BSs selected for echo signal reception will also hear the crosstalk from other DFRC BSs. This potentially degrades the sensing performance of the ISAC system.
        \item \textbf{Multiuser Interference}: MUI is one of the main types of interference in conventional wireless communications, and limits the sum rate of multiuser systems. Unfortunately, the nature of ISAC systems can further aggravate this situation. On the one hand, a high-energy ISAC signal may dramatically increase the MUI for some CUs. On the other hand, due to the existence of STs, the information signal of one CU is likely to be reflected to unintended users, leading to additional MUI compared to conventional wireless communication systems.
\end{itemize}
\section{Interference Mitigation Techniques and Corresponding Optimization Problems}
\begin{table*}[t]
  \centering
  \caption{Comparison of different interference mitigation techniques.}
    \begin{tabular}{|c|c|c|c|c|c|c|}
     \hline
     \textbf{Technique} & \textbf{Domain} & \textbf{Usage} & 
     \textbf{Limitations} & \textbf{Complexity}\\
     \hline
     \textbf{CMT} & Space & SI, MI, crosstalk & Requires signaling exchange and synchronization & High \\
     \hline
     \textbf{IA} & Space & MI, MUI & Requires accurate CSI of both systems & Moderate \\ \hline
     \textbf{HD-BF} & Space & Clutter & Requires prior knowledge of STs & Moderate\\ 
     \hline 
     \textbf{TS} & Time & MI, MUI & Causes discontinuous service & Low\\ \hline
     \textbf{SA} & Frequency & MI, MUI, crosstalk & Lowers data rate & Low\\ 
     \hline 
    \end{tabular}%
  \label{tab:table1}%
\end{table*}
\begin{table*}[t]
		\caption{Comparison of different interference mitigation techniques from an optimization perspective.}
		\label{tab:table2}
		\newcommand{\tabincell}[2]{\begin{tabular}{@{}#1@{}}#2\end{tabular}}
		\centering
		\begin{tabular}{|c|c|c|c|c|}\hline
		\textbf{Technique} & \textbf{Problem type} & \textbf{Convex} & \textbf{Technical challenge} & \textbf{Method}\\
			\hline
   	\textbf{CMT}	   & CPP & No & Cardinality constraint & GBD \cite{geoffrion1972generalized}, big-M method, SCA\\
			\hline
		\textbf{IA}		   & RCOP & No & Rank equality constraint & ALP-based method \cite{escalante2011alternating}\\
                \hline
	\textbf{HD-BF}	   & SDP & Yes & None & Convex optimization methods\\
			\hline
        \textbf{TS}		   & BLP & No & Bilinear term & AO \cite{bezdek2002some}, BT
        \\ 			\hline
	\textbf{SA}	       & BIP & No & Binary constraint & BnB, LPR, penalty method \cite{ruszczynski2011nonlinear}\\ \hline
		\end{tabular}
\end{table*}
In this section, we present several advanced interference mitigation techniques for network-level ISAC systems and discuss their advantages and disadvantages. Subsequently, for each approach, we identify the design challenges and propose corresponding optimization methods to effectively mitigate the interference. The characteristics of the proposed interference mitigation techniques and the corresponding optimization problems are summarized in Table \ref{tab:table1} and Table \ref{tab:table2}, respectively. Readers who are interested in mitigation techniques for different interferences may focus on Table \ref{tab:table1}. Readers who are more interested in the corresponding optimization method can refer to Table \ref{tab:table2}. Note that each technique is able to mitigate several types of interferences by exploiting the resources available in a certain domain. Moreover, more advanced techniques lead to higher design requirements and implementation complexity, while relatively straightforward techniques enable simpler implementation at the cost of system performance. 
\subsection{Coordinated Multipoint Transmission}
\noindent\textbf{Technique description:} Coordinated multipoint transmission (CMT) is an effective means to mitigate SI, MI, and crosstalk by coordinating the transmissions of multiple nodes such as APs, BSs, and remote radio heads (RRHs). Specifically, all nodes are synchronized and grouped into a cooperative cluster, which shares the spectrum and CSI. With CMT, we can extend the conventional monostatic radar architecture to a more general bistatic/multistatic radar architecture \cite{8999605}. According to the network geometry and CSI of the system, each node can be assigned to carry out different tasks. This not only allows separating sensing and communication in the space domain but also isolates echo signal reception from sensing/communication signal transmission, which helps reduce MI and SI. Moreover, by jointly designing the resource allocation of the whole system, the crosstalk between different nodes can be appropriately managed. Yet, to effectively utilize CMT, accurate CSI of the whole system needs to be obtained, which is a challenging task in practice. Moreover, the synchronization between different nodes requires a large amount of signaling exchange.
\par
\noindent\textbf{Optimization problem:} The most critical issue for CMT is how to exploit the wireless resources of the entire network to provide satisfactory ISAC services. To meet prescribed quality-of-service (QoS) requirements for sensing and communication, the node assignment policy in CMT should be jointly designed with the beamforming policy. Moreover, due to the limited spectrum, computation, and storage resources, a given node may not be able to simultaneously serve all CUs and collect the echo signals from all STs. Instead, only a subset of CUs or STs can be served by one node and the cardinality of the subset should be determined based on the network conditions and requirements. As a result, there is a non-convex cardinality constraint for each node, leading to a combinatorial programming problem (CPP). To optimally solve the resulting non-convex problem, one may resort to the application of generalized Bender's decomposition (GBD) theory \cite{geoffrion1972generalized}. On the other hand, the big-M method and successive convex approximation (SCA) are two promising enablers for finding suboptimal solutions to the considered problem.
\subsection{Interference Alignment}
\noindent\textbf{Technique description:} When the nodes of the ISAC network are equipped with multiple antennas, we can employ multiple-input multiple-output (MIMO)-based interference alignment (IA) to manage the MI and MUI in the space domain \cite{7442513}. Given the CSI of the entire network, we can design the beamforming policy such that the sensing beam lies in the null spaces of the channels of all the CUs. Going one step further, the expected signal for each CU can be retrieved by constructing a decoding matrix that nullifies the interference. By doing so, we can completely mitigate MI and MUI. However, IA can be performance-conservative and inefficient when the number of antennas is less than the number of CUs and STs. Moreover, the effectiveness of IA heavily depends on the accuracy of the CSI. As a result, a certain portion of radio resources must be dedicated to frequent CSI updates and signaling exchanges.
\par
\noindent\textbf{Optimization problem:} To facilitate high-quality ISAC, the IA-based beamforming policy should satisfy a set of non-convex equality constraints. First, for the communication system, the interference signal (i.e., the sensing signal) should be orthogonal to all CUs' channels, leading to a set of equality constraints. Second, we need to ensure that the information signal for a specific CU lies in the null space of the channels of the other CUs, leading again to a set of equality constraints. Third, to guarantee that the information signal can be reliably received by the intended CU, the received signal at each CU should have non-zero dimensions, resulting in a rank equality constraint. As a result, a non-convex rank-constrained optimization problem (RCOP) has to be solved when employing IA in ISAC systems. Unfortunately, it is very challenging to optimally and efficiently solve such an RCOP. As a compromise, low-complexity suboptimal methods, e.g., the alternating projection (ALP)-based method \cite{escalante2011alternating}, have been developed. The fundamental idea behind the ALP-based method is to replace the rank constraint with a more tractable convex constraint.
\subsection{Highly-Directional Beamforming}
\noindent\textbf{Technique description:} Clutter is caused by reflections of the ISAC signal from non-relevant targets or the ground \cite{skolnik2008radar}. A promising means to mitigate clutter is to employ highly-directional beam patterns (HD-BPs). Specifically, given a direction of interest, an ideal flat-top beam pattern can be pre-designed and generated. The ideal beam pattern represents the power allocation of the beam in each azimuth angle, where only the direction of interest is illuminated with the narrow main lobe of the sensing beam. Yet, in practice, it is very challenging to generate such an ideal beam. Instead, by approximating the ideal beam, practical ISAC signals with side lobes are synthesized to illuminate the direction of interest. 
\par
\noindent\textbf{Optimization problem:} To facilitate the application of HD-BP, the mismatch between the actual beam pattern and the ideal beam pattern needs to be restricted, leading to a convex constraint. The resulting semidefinite programming (SDP) problem can be optimally and efficiently solved.
\subsection{Task Scheduling}
\noindent\textbf{Technique description:} MI and MUI can be effectively eliminated by scheduling sensing and communication tasks orthogonally in the time domain. Specifically, by applying task scheduling (TS), we can divide a given frame into a series of time slots, and schedule a subset of nodes to serve CUs or sense desired targets in different time slots. The length of each time slot can be fixed or designed frame by frame according to the specific requirements of each task. We note that the order of tasks is usually determined by the specific application scenario. Moreover, the duration of each task should be determined according to the specific QoS requirements. For example, if only coarse knowledge of the ST is needed, while the QoS requirement for communication is high, then a longer communication period is preferred. By employing task scheduling, we can effectively circumvent MI and MUI. Moreover, by scheduling the tasks across different DFRC BSs, BS crosstalk can also be suppressed. However, task scheduling inevitably leads to discontinuity in information transmission, which may reduce the average per-frame achievable rate. Also, the seamless support of time-sensitive tasks such as target tracking may not be possible. 
\par
\noindent\textbf{Optimization problem:} TS usually leads to variable coupling, which is challenging for optimization algorithm design. In particular, the duration of each task (associated with a time allocation optimization variable) is typically multiplied by other variables (e.g., the transmit beamforming vector or transmit power), leading to a non-convex bilinear programming (BLP) problem. In the literature, the alternating optimization (AO) approach is widely adopted to overcome the multiplication of optimization variables \cite{bezdek2002some}. Also, the bilinear transformation (BT) approach can be employed to handle the multiplication of optimization variables. In particular, BT sidesteps the variable coupling issue by treating the multiplication of the time allocation variables and the coupled variables as a new entirety. To ensure the equivalence of such BT, two additional constraints, i.e., a difference of convex functions constraint and a linear constraint, have to be imposed on the original optimization problem.
\subsection{Subcarrier Assignment}
\noindent\textbf{Technique description:} In addition to exploiting the time domain, subcarrier assignment (SA) can be utilized to perform different tasks at different nodes in a frequency-orthogonal manner, which facilitates the mitigation of MI, MUI, and crosstalk. In particular, SA allows the division of the available spectrum into non-overlapping frequency bands. The division can be pre-determined in an offline manner or dynamically adapted to the real-time demand in an online manner. Based on the optimized SA policy, the DFRC BS and CUs can acquire their desired information from the associated frequency bands. In fact, by employing SA, we can completely eliminate a large amount of interference in ISAC systems, including MI, crosstalk, and MUI. However, applying SA in an ISAC system may reduce the sum rate of the system, as a portion of the spectrum has to be allocated for sensing.
\par
\noindent\textbf{Optimization problem:} The SA in ISAC systems can be formulated as a binary integer programming (BIP) problem. Although the feasible set of BIP problems is naturally non-convex, they can be optimally solved by well-established enumeration algorithms such as branch-and-bound (BnB) algorithms. Alternatively, BIP can be suboptimally and efficiently tackled using low-complexity algorithms such as linear programming relaxation (LPR) and penalty-based methods. 
\section{Case Studies}
In this section, we investigate two network-level ISAC systems to validate the effectiveness of different interference mitigation techniques. In particular, we first focus on the design of a network-based ISAC system supported by CMT and HD-BF. Then, by exploiting TS and CMT, an efficient interference mitigation design of a distributed antenna-assisted ISAC system is investigated. For both cases, we aim to minimize the total transmit power and the proposed optimization framework can also be applied to solve other optimization problems, including system sum rate maximization and Cram\'{e}r-Rao lower bound (CRLB) minimization. In the simulation, we focus on the typical scenario where a small set of static CUs and STs are uniformly and randomly distributed. More comprehensive simulation scenarios, e.g., mobile environment and dense environment, will be considered in future works.
\subsection{Coordinated Cellular Network-Based ISAC}
\begin{figure}[t]
\centering
\includegraphics[width=3.4in]{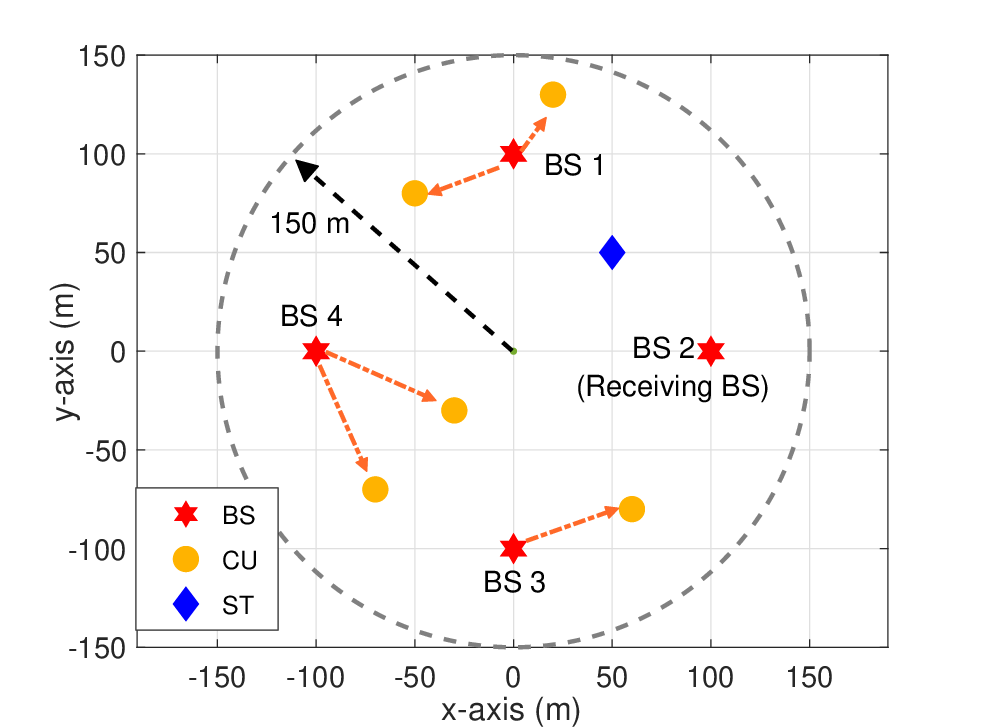}
\caption{Illustration of the considered ISAC system with four BSs, five CUs, and one ST.}
\label{setup}
\end{figure}
We consider a coordinated cellular network-based ISAC system with multiple DFRC BSs, multiple CUs, and one ST. The DFRC BSs are equipped with multiple antennas while both the CUs and the ST are assumed to be single-antenna devices. To evaluate the performance upper bound of the coordinated cellular network, we assume that all the BSs are perfectly synchronized. We jointly exploit the wireless resources of all coordinated base stations in both power and space domains for interference management. In this case, we propose to select one BS for collecting echo signals, while the other BSs collaborate to serve as transmitters. This case study aims at minimizing the total transmit power to achieve satisfactory communication and sensing services, by jointly optimizing the BS selection, user association, and beamforming policy with the AO method. The coupling of the optimization variables and the binary constraint are tackled by employing the big-M method and penalty method, respectively. 
\par
Fig. \ref{setup} shows the BS selection and user association results for a system setting with a circular service area (radius$=150$ m). To ensure high-quality ISAC, CRLB and signal-to-interference-plus noise ratio (SINR) are adopted as the performance metrics for sensing and communication, respectively. Here, the CRLB of each ST and the SINR of each CU are set to $1$ and $8$ dB, respectively. As can be observed in Fig. \ref{setup}, BS2 is selected as the echo signal receiver. Meanwhile, the other BSs are associated with different CUs, as indicated by the orange dashed arrows. 
\par
\begin{figure}[t]
\centering
\includegraphics[width=3.4in]{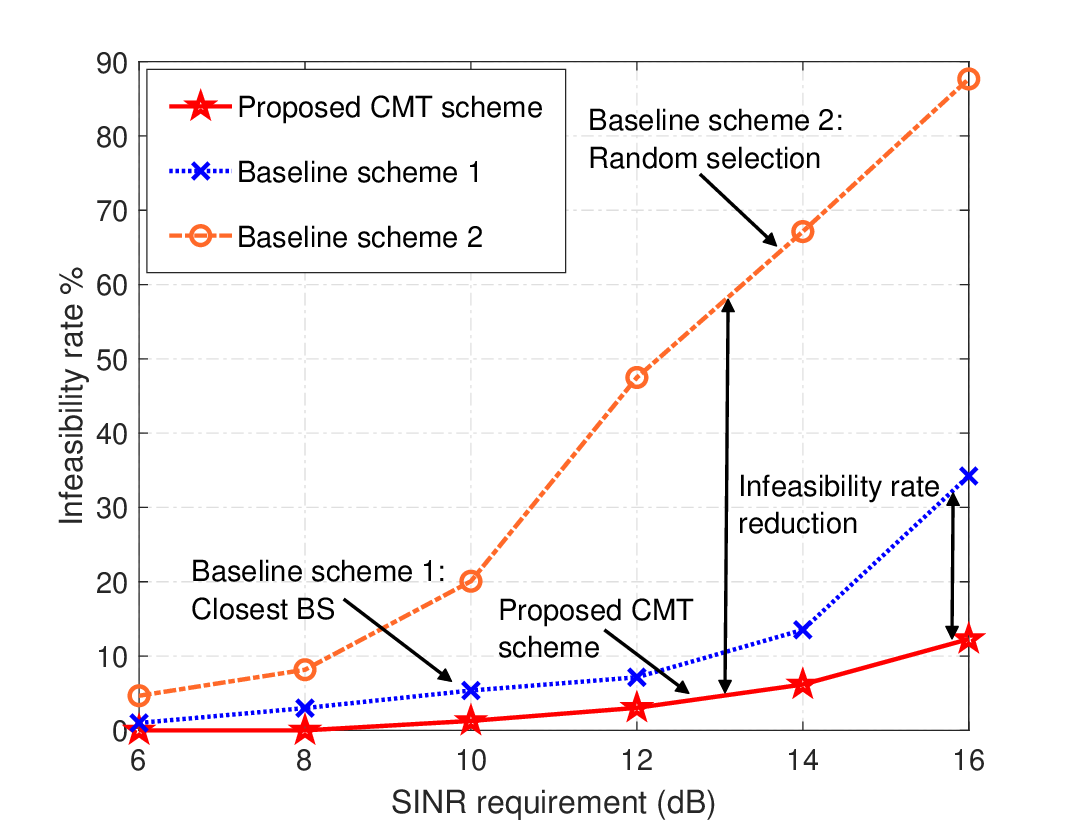}
\caption{Infeasibility rate versus minimum SINR requirement.}
\label{figure:infeasibility_rate_sinr}
\end{figure}

In Fig. \ref {figure:infeasibility_rate_sinr}, we show the infeasibility rate versus the SINR requirements of the CUs. Specifically, for each considered SINR, we randomly generate $100$ different system setups. Then, for each system setup, we solve the corresponding optimization problem and count the number of infeasible solutions caused by interference. Here, the horizontal and vertical coordinates denote the SINR requirement of the communication users and the infeasibility rate of solving the $100$ optimization problems with different system setups, respectively. To better evaluate the effectiveness of the proposed scheme, we also consider three baseline schemes for comparison. For baseline scheme 1, to avoid SI, we use the bistatic sensing architecture where BS1 and BS2 are selected as the transmitting BS and the receiving BS, respectively. Baseline scheme 2 employs a random receiving BS selection and CU assignment policy. As for baseline scheme 3, rather than coordinated ISAC, the monostatic radar architecture is adopted and the nearest BS is used to sense the target while the other BSs perform information transmission for all CUs. We can observe from Fig. \ref {figure:infeasibility_rate_sinr} that as the SINR requirement becomes more rigorous, the infeasibility rate of the proposed scheme and the three baseline schemes monotonically increase. This is because, to satisfy a larger SINR of the communication users, the system has to consume more transmit power on information transmission. Once the required transmit power is larger than the power budget of the system, the corresponding optimization problem becomes infeasible. Moreover, it can be observed that there is a prominent gap between the proposed scheme and the three baseline schemes. This validates the benefits of the proposed scheme in effectively mitigating different types of interference, including SI, MI, and MUI, compared to the three baseline schemes.
\subsection{Distributed Antenna-Based ISAC}
Besides coordinated cellular networks, ISAC can also be implemented on a distributed antenna network (DAN) which comprises a central processor (CP) and multiple RRHs. Compared to coordinated cellular networks, DANs have the following advantages. First, each RRH is connected with the CP via an individual fronthaul link, which makes signaling and synchronization more efficient. Second, as the core unit of the DAN, the CP has sufficient computation resources to carry out all computationally intensive tasks in the ISAC system. Next, we investigate the design of a DAN-based ISAC system. In this case, we propose to apply TS to partition each ISAC frame into a communication phase and a sensing phase. In the communication phase, the CP optimizes the beamforming policy and forwards the communication data as well as the resource allocation control signals to the RRHs via the fronthaul links. Also, we propose to pre-select a dedicated RRH for each ST to facilitate efficient target sensing based on HD-BF. As such, in the sensing phase, the CP optimizes the sensing signal for each RRH-ST pair and delivers it to RRHs via fronthaul links. Then, adopting a pulse radar mechanism, all RRHs concurrently employ HD-BF to illuminate their associated ST before switching to the listening mode to receive the echoes. The collected echo signals are conveyed to the CP via the fronthaul link for sensing information extraction. For a given time horizon and a pre-designed RRH-target assignment policy, we jointly optimize the time allocation and beamforming policy to minimize the total energy consumption, subject to fronthaul link capacity constraints and sensing and communication performance constraints. The minimum echo power strength and the achievable rate are adopted as the performance metrics for sensing and communication, respectively. The formulated non-convex BLP problem is tackled by capitalizing on AO and SCA approaches.
\par
\begin{figure}[t]
\centering\includegraphics[width=3.4in]{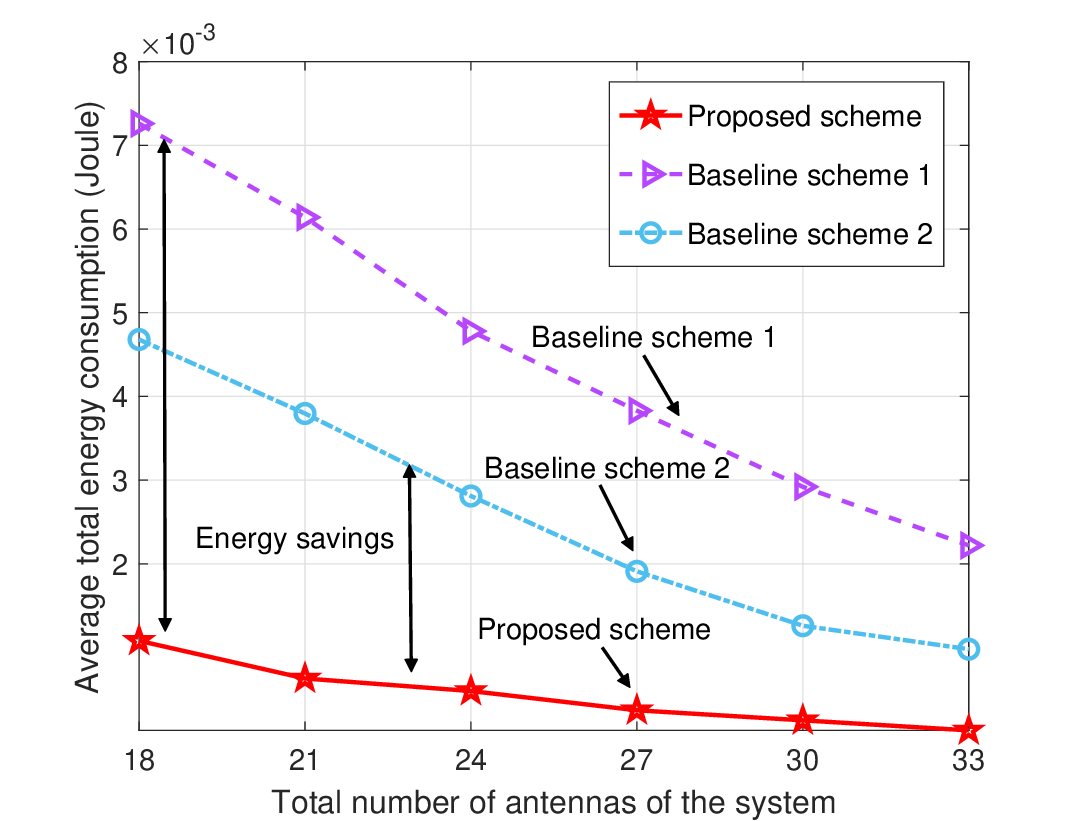} 
\caption{Average total energy consumption during an ISAC frame (dBm) versus the total number of transmit antennas for different schemes.} \label{fig:Distri_Ant_ISAC_Energy}
\end{figure}

Fig. \ref{fig:Distri_Ant_ISAC_Energy} shows the average energy consumption in an ISAC frame versus the total number of transmit antennas with the minimum required achievable rate of $2$ bits/s/Hz and the minimum required echo power of $-90$ dBm. Here, the horizontal and vertical coordinates denote the total number of antennas of all RRHs and the average total energy consumption of the system in the given time horizon, respectively. We consider two baseline schemes for comparison. For baseline scheme 1, a BS with co-located transmit antennas is adopted to perform ISAC. In this case, a multi-beam pattern is employed to concurrently sense all STs. As for baseline scheme 2, the time horizon is equally divided between the communication phase and the sensing phase. As can be observed from Fig. \ref{fig:Distri_Ant_ISAC_Energy}, the average total energy consumption decreases with the total number of antennas. This is due to the fact that additional antennas can be exploited to perform more accurate HD-BF. Moreover, compared to the proposed scheme, both baseline schemes consume more energy. This is because the co-located antenna architecture employed in baseline scheme 1 does not offer spatial macro-diversity to combat MI, MUI, and clutter, while the power gain of baseline scheme 2 is limited by the fixed duration of the communication and sensing phases. On the contrary, the proposed scheme can prominently mitigate SI, MI, and MUI, and thus less total energy is needed to satisfy both sensing and communication quality-of-service requirements. The above observations validate the effectiveness of the joint time and signal optimization for the proposed DAN-based ISAC system.
\section{Conclusions and Future Research Directions}
In this article, we provided a comprehensive overview of the typical interference in network-level ISAC systems including SI, MI, clutter, crosstalk, and MUI. To effectively address these different types of interference, we introduced several efficient interference mitigation techniques. Subsequently, for each technique, we explained the resulting optimization problems and proposed several promising optimization methods. To pave the way towards fully unleashing the potential of network-level ISAC, additional unremitting efforts on the following open research questions have to be made.
\par
\textbf{Comprehensive interference mitigation framework design:} Although different techniques can be used to combat one or multiple types of interference, it is difficult to completely suppress all types of interference with one technique. Hence, multiple techniques have to be jointly exploited and a comprehensive interference mitigation framework that can simultaneously overcome various types of interference has to be developed. This inevitably leads to more complex resource allocation optimization problems, thus requiring more advanced optimization algorithms.
\par
\textbf{Intelligent reflecting surfaces-enhanced ISAC:} As a revolutionary technique in 6G networks, intelligent reflecting surfaces (IRSs) can be flexibly integrated into wireless networks to create favorable radio propagation environments. In fact, by smartly programming the phase shift configurations of the IRS, we can improve the received signal power in the network-level ISAC network to facilitate high-quality ISAC. Yet, the combination of IRS and ISAC also brings new challenges such as IRS-induced cascaded channel estimation, joint IRS and BS resource allocation design, and the synchronization between IRS and BS. Such issues must be carefully addressed to unlock the full potential of IRS-enhanced network-level ISAC. 
\par
\textbf{Interference mitigation with CSI uncertainty:} Due to the integration of radar and information transmission with shared hardware equipment, ISAC systems usually suffer from severe CSI estimation errors. Moreover, for network-level ISAC, it is in general challenging to obtain perfect and up-to-date CSI of the whole system by employing existing channel estimation schemes. As a result, CSI uncertainties have to be taken into account when designing interference mitigation algorithms to ensure robust ISAC. 
\par
\textbf{Deep learning-enabled interference mitigation:} Although a plethora of optimization techniques have been reported for mitigating various types of interference in ISAC systems, the resulting computational complexity may be unaffordable for network-level ISAC. To this end, deep learning-based techniques are promising solutions to facilitate low-complexity interference mitigation algorithms for practical ISAC systems. In particular, based on the existing mathematical models, model-driven techniques can be employed to exploit the characteristics of radar and communication systems, which potentially reduces the required huge amount of training data while facilitating low-complexity resource allocation design for practical ISAC systems.


%

%

%
%

\ifCLASSOPTIONcaptionsoff
  \newpage
\fi



%
%
%
\bibliographystyle{IEEEtran}
\bibliography{Reference_List}

%

%
%
%




\end{document}